\documentstyle[multicol,aps]{revtex}

\input{epsf}
\epsfverbosetrue \epsfclipon

\newcommand{\putfigxsz}[4]{
   \begin{figure}\begin{center}\mbox{\epsfxsize #4
   \epsffile{#1}}
   \end{center}
   \end{figure}
   }

\begin{document}

\draft
\title{Universal behaviour in the floating up of quantum Hall extended states as B $\rightarrow$ 0}
\author{C. E. Yasin, M. Y. Simmons, A. R. Hamilton, N. E. Lumpkin, and R.
G. Clark}
\address{School of Physics, University of New South Wales, Sydney
2052, Australia}
\author{L. N. Pfeiffer and K. W. West}
\address{Bell Laboratories, Lucent
Technologies, Murray Hills NJ 07974, USA}

\date{\today}
\maketitle
\begin{abstract}

We investigate the relationship between the quantum Hall extended
states and the apparent $B=0$ `metal'-insulator transition in
extremely low density, high quality two-dimensional n-GaAs systems
($\mu_{peak} \sim 2 \times 10^7$ cm$^2$V$^{-1}$s$^{-1}$). The
combination of small effective mass and high quality allow us to
resolve the continued floating up in energy of the extended states
down to very low magnetic fields, B=0.015T, despite an apparent
`metal'-insulator transition at B = 0. We present a modified
global phase diagram which brings together conflicting data in the
literature from different material systems, and reconciles the
differences as to the fate of the extended states as $B
\rightarrow$ 0.

\end{abstract}


\begin{multicols}{2}

There has been intense interest in the unexpected metallic
behaviour observed in dilute two dimensional (2D) electron and
hole systems. While the one parameter scaling theory of
localization \cite{Scaling} predicts that in the absence of a
magnetic field all 2D systems are insulating at $T$ = 0, an
anomalously strong `metallic'-like behaviour ($dR/dT >$ 0) has
been observed in various high quality, strongly interacting (large
$r_s$) systems over the last few years
\cite{2DMIT,2DMIT-Popovic,2DMIT-Pudalov,2DMIT-pGaAs4,2DMIT-pGaAs,2DMIT-pGaAs1,2DMIT-pGaAs2,2DMIT-pGaAs3,2DMIT-pSiGe,2DMIT-nAlAs,2DMIT-nGaAs}.
Although this temperature dependence suggests the possibility of a
`metallic' ground state, recent experiments
\cite{Corrections,Corrections1,Prus} have shown that even in the
`metallic' phase there are logarithmic localising corrections
which appear to drive the conductivity to zero as $T \rightarrow$
0. If these corrections are not overcome by strong interaction
effects \cite{Finkelstein} the results indicate that the
`metallic'-like behaviour of $dR/dT$ may only be a finite
temperature effect and the system is insulating at $T$=0. Despite
numerous experimental and theoretical proposals the questions of
what causes `metallic' behaviour at finite temperatures, and if a
2D system can be a true metal at $T$ = 0, remain a subject of
controversy.

Experimentally it is not easy to attain temperatures below a few
tens of millikelvin, making the detection of logarithmic
insulating behaviour in high quality samples where `metallic'
behaviour is observed difficult. Another way to determine the
ground state is to relate what happens to the extended states in
the quantum Hall regime at finite magnetic fields to the apparent
`metal'-insulator transition (MIT) at $B$ = 0 \cite{Hanein}. If
there are extended states below the Fermi energy at $B$ =0 then
the system will be metallic at $T$ = 0, otherwise it is an
insulator. Since the one parameter scaling theory \cite{Scaling}
shows that non-interacting 2D systems are always localised,
Khmelnitskii \cite{Floating} and Laughlin \cite{Floating1} argued
that the quantum Hall extended states `float up' in energy as $B
\rightarrow$ 0, so that only localised states remain below the
Fermi energy. This floating up has been observed in disordered
n-GaAs and p-Ge systems\cite{Glozman,Hilke} that are insulating at
$B$ = 0. However, recent studies of high quality p-GaAs systems,
which exhibit `metallic' behaviour at $B$ = 0, indicate that the
Landau levels remain at a finite energy as $B \rightarrow$0
\cite{Hanein,Dultz}, suggesting a link between the $B$ = 0
`metallic' behaviour and the quantum Hall extended states.

In this paper we track the evolution of the quantum Hall extended
states as $B \rightarrow$ 0 in extremely high quality GaAs
electron systems that exhibit an apparent MIT at $B$ = 0. Two
different samples from undoped GaAs/AlGaAs heterostructures, where
carriers are induced with a gate bias \cite{Kane}, were studied.
These samples are amongst the highest quality systems in which the
2D MIT has been observed ($\mu_{peak}\sim 2\times 10^7$
cm$^2$V$^{-1}$s$^{-1}$). The main difference between the samples
is the lowest carrier density reached: $5 \times 10^9$ cm$^{-2}$
in sample A and $3 \times 10^9$ cm$^{-2}$ in sample B. Most
significantly the electron effective mass ($0.067 m_e$) is low
compared to other semiconductor systems, and since the Landau
level separation scales as $eB/m^*$ we are able to follow the
Landau levels to very low magnetic fields -- precisely where
`floating up' should be most pronounced. 

\begin{figure}[tbph]
\putfigxsz{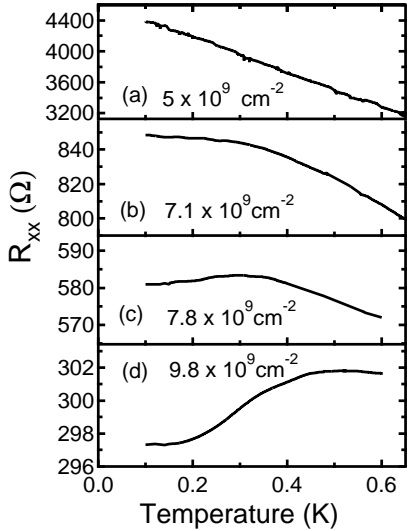}{lab}{cap}{6cm} \caption{Longitudinal
resistance of sample A as a function of temperature at $B$ = 0 for
the carrier densities indicated.} \label{2DMIT} \vspace{0.1in}
\end{figure}

Figure \ref{2DMIT} shows the resistance of sample A as a function
of temperature at $B$ = 0 for different carrier densities on both
sides of the `metal'-insulator transition. At low densities $n_s =
(5-7.1)\times 10^9$ cm$^{-2}$, the sample shows insulating
behaviour where $R_{xx}$ increases as $T\rightarrow$0. As the
density is increased to $7.8 \times 10^9$ cm$^{-2}$ the resistance
exhibits non-monotonic behaviour, with insulating-like behaviour
($dR/dT < 0$) at high temperatures and `metallic'-like behaviour
($dR/dT > 0$) at lower temperatures (Fig. \ref{2DMIT}(c)). The
change in sign of $dR/dT$ occurs at $T/T_F \sim$ 0.1, where $T_F$
is the Fermi temperature. This non-monotonic behaviour is due to
the system becoming non-degenerate \cite{Das-Sarma} as has been
observed in other material systems when $T/T_F \sim$ 0.1 - 0.7
\cite{2DMIT-pGaAs,2DMIT-pGaAs1,2DMIT-pGaAs2,2DMIT-pGaAs3,2DMIT-nAlAs,2DMIT-nGaAs}.
As the density is increased further, to $9.8 \times 10^{9}$
cm$^{-2}$, the sample shows `metallic' behaviour over the complete
measurement range with the resistance dropping as $T \rightarrow$
0 (Fig. \ref{2DMIT} (d)). The `metallic'-like drop in resistance
in this sample ($\sim 5\%$) is consistent with other studies where
a decrease of anything between $3\%$ \cite{2DMIT-nGaAs} and a
factor of eight \cite{2DMIT} has been observed.

\begin{figure}[tbph]
\putfigxsz{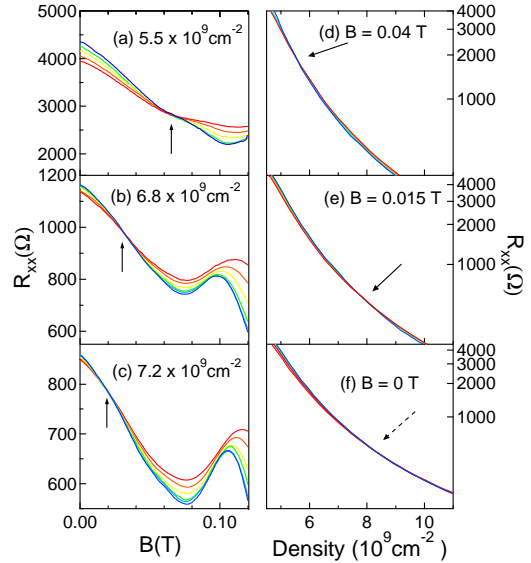}{lab}{cap}{7cm} \caption{Temperature
dependence of the magnetoresistance at 100, 150, 200, 250 and
300mK for sample A: (a-c) as a function of magnetic field for
different densities and (d-f) as a function of density for
different magnetic fields. The arrows mark the transition from
insulator to quantum Hall liquid.} \label{xpoints} \vspace{0.1in}
\end{figure}

Having demonstrated the existence of an apparent `metal'-insulator
transition at $B$ = 0, we now turn to examine the evolution of the
extended states in the quantum Hall regime as $B \rightarrow$ 0.
Measurements of the magnetoresistance $R_{xx}$  at different
temperatures are used to identify the quantum Hall
liquid-insulator transitions, using two well-established methods
\cite{Hanein,Glozman}. This approach has the advantage that the
transitions between quantum Hall and insulating states define a
quantum critical point and provides an independent means of
determining the ground state of the system at $B$ = 0. In the
first method, $R_{xx} (B)$ is measured at a fixed carrier density
for different temperatures. Typical traces are shown in Figs.
\ref{xpoints}(a-c) for densities that are insulating at $B$ = 0.
The arrows mark the temperature independent points, corresponding
to the centre of the Landau levels which separate the insulating
state ($dR/dT <$ 0) at low $B$ from the quantum Hall liquid
($dR/dT >$ 0) at intermediate $B$. In the second technique, the
density dependence of $R_{xx}$ is measured for different
temperatures at a fixed $B$ (Figs. \ref{xpoints}(d-f)). At `high'
magnetic fields ($B \ge$ 0.04 T), the resistance depends strongly
on temperature, yielding clear transitions from insulating to
`metallic' behaviour (indicated by the arrow in Fig.
\ref{xpoints}(d)). However, at lower magnetic fields (Fig.
\ref{xpoints}(e)), the separation between the Landau levels is
much smaller. This makes it difficult to identify the transitions,
particularly for these high quality n-GaAs samples where only a
small change in $R_{xx}$ is observed over the measurement
temperature range. While it is tempting to use this method to go
to even lower fields and make a direct link to the apparent MIT at
$B$ = 0 (Fig. \ref{xpoints}(f)) \cite{note}, it is not
experimentally possible to track the Landau levels below 0.015 T.
This is due to the fact that the Landau level separation at these
small fields becomes comparable to the thermal broadening (at $B$
= 0.01 T, $\hbar\omega_c \sim 4kT$ at $T$ = 50 mK).

\begin{figure}[tbph]
\putfigxsz{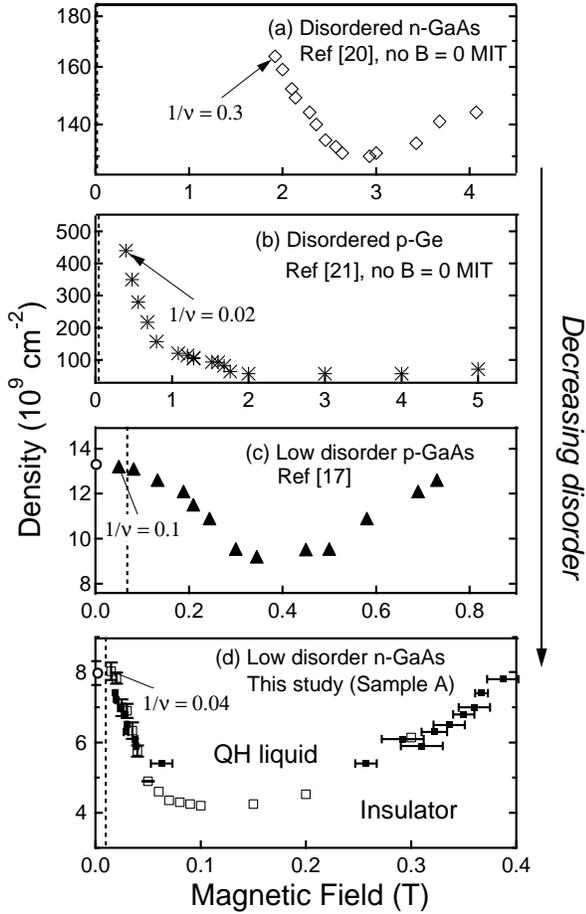}{lab}{cap}{8cm}\caption{Position of the
extended states as a function of density and magnetic field, from
studies in different material systems. The open and closed squares
in (d) mark the position of the extended states with corresponding
error bars obtained by the two different methods discussed in the
text. The dotted lines in (a-d) mark the magnetic field below
which thermal broadening becomes significant, $\hbar\omega_c/4k <
50$ mK.} \label{floating}
\end{figure}

To track the evolution of the Landau levels in magnetic field, we
have repeated these measurements at many different carrier
densities and magnetic fields, and plot the position of the
$T$-independent points in Fig. \ref{floating}(d). Solid symbols
mark points determined from the fixed density measurements while
open symbols are obtained from fixed field measurements. Good
agreement is obtained between these two techniques, and the data
in Fig. \ref{floating}(d) clearly shows that for $B <$ 0.05 T the
extended states float up rapidly as $B \rightarrow$ 0. The power
of this technique is that these transition points are
$T$-independent and should therefore persist down to $T$ = 0 since
they mark a quantum phase transition between quantum Hall liquid
and insulating states. However at extremely low magnetic fields it
is no longer possible to resolve the Landau levels and this
technique is not valid. In particular there are no Landau levels
at $B$=0, and whilst there is an apparent transition in Fig.
\ref{xpoints}(f) it is not known if it persists to $T$ =
0~\cite{Corrections,Corrections1,Prus}. It is therefore crucial to
track the quantum Hall transitions to as low a magnetic field as
possible to shed light on the $B$ = 0 ground state.

In these experiments we have tracked the extended states down to
the lowest fields that they can be resolved ($B$ = 0.015 T), and
find that they float up continuously with no signs of saturation
as $B \rightarrow$ 0. Furthermore the slope $dn_s/dB$ of the
Landau level trajectory at these small fields ($0.015 < B < 0.05$
T) is more that an order of magnitude larger than previous studies
in p-GaAs \cite{Hanein}. The observation that the extended states
start to float up so rapidly as $B \rightarrow$ 0 makes it
difficult to relate the quantum Hall extended states in finite
magnetic fields to the apparent $B$ = 0 MIT at $n_s^c = (8.0\pm
0.25) \times 10^9$ cm$^{-2}$ \cite{note}, since this would require
a sharp change in the trajectory of the extended states below
0.015T.

To compare these results with previous studies we show in Figs.
\ref{floating} (a-c) similar data from disordered n-GaAs
\cite{Glozman}, disordered p-Ge \cite{Hilke}, and high quality
p-GaAs \cite{Hanein}. In both of the disordered systems no
apparent `metallic' behaviour is observed at $B$ = 0, and the
Landau levels float up rapidly as $B \rightarrow$ 0. However in
p-GaAs, where  `metallic'-like behaviour is observed at B=0, the
Landau levels start to float up, but then appear to saturate to a
finite density as $B\rightarrow$0. This observation has led to the
suggestion that there is a relation between the quantum Hall
effect and the `metallic' behaviour observed at $B$ = 0 in these
systems \cite{Hanein}.

To resolve the discrepancy between the various experiments as to
the fate of the extended state as $B\rightarrow$0 we must consider
both the effects of disorder and finite temperature. It is known
that the Landau levels start to float up when $\omega_c\tau = \mu
B \sim$ 1 \cite{Floating,Floating1}. In disordered n-GaAs (Fig.
\ref{floating}(a)) this floating up occurs at relatively high
magnetic fields $B \sim $ 2.8 T such that the small effective mass
makes it easy to resolve the Landau level separation $\hbar e
B/m^{*}$. For disordered p-Ge in Fig. 3(b) the effective mass is
higher, but the strong disorder still means the floating up occurs
at large fields, $B \sim$ 2 T, where the Landau levels are easily
resolved.

The floating up becomes more difficult to observe experimentally
in high quality samples (where the apparent $B=0$ metallic
behaviour occurs) since we require very low temperatures, $kT \ll
\hbar eB/m^*$, to avoid thermal smearing\cite{DasSarma}. In the
high quality p-GaAs of Fig. 3(c) the low disorder means that the
levels only start to float up at low magnetic fields, $B$=0.4 T,
where due to the large hole mass the Landau level spacings are
already inherently small and thermal broadening becomes
significant. The vertical dotted lines in Fig. \ref{floating}(a-d)
mark the magnetic fields below which it is not possible to track
the Landau levels, because the level separation becomes comparable
to the thermal broadening at 50mK. This line clearly shows that
resolving the Landau levels below $B \sim 0.08$ T is problematic
for the high quality p-GaAs sample in Fig. \ref{floating}(c).
However for low disorder n-GaAs the small electron mass makes it
possible to safely track the Landau levels to much smaller
magnetic fields, as can be seen in Fig. \ref{floating}(d). Here
we can reach magnetic fields as low as $B \sim$ 0.015 T before
thermal broadening sets in, and down to these low magnetic fields
there is no sign of saturation. For completeness we have also
marked the apparent $B$ = 0 transition, but re-iterate that
unlike the quantum Hall transitions at finite $B$, it is not
known if this transition persists to $T$ = 0. Significantly, this
is the first time that continuous floating up of the Landau levels
has been observed in high quality samples that exhibit `metallic'
behaviour at $B=0$.

\begin{figure}[tbph]
\putfigxsz{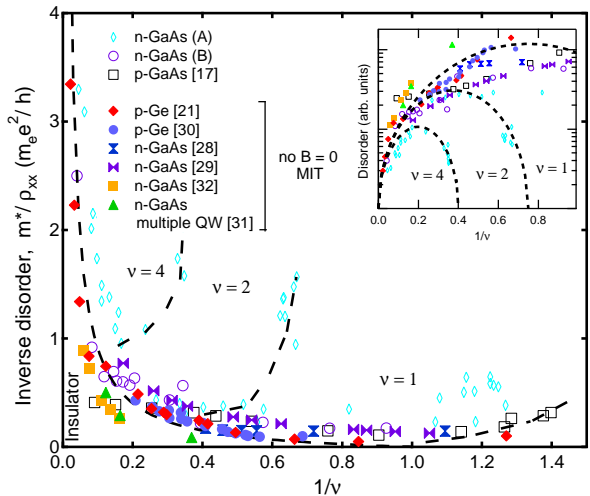}{lab}{cap}{8.5cm} \caption{A modified form of
the global phase diagram (normalised for the effective mass),
plotted in the $1/\nu$ versus inverse disorder plane. The inset
shows the schematic global phase diagram of Ref. [25] with the
same experimental data.} \label{phase-diagram} \vspace{0.1in}
\end{figure}

Although both p- and n-GaAs show floating up of the Landau
levels, an apparent discrepancy still remains. In Fig.
\ref{floating}(c) it might appear that at intermediate fields
($B$ = 0.2 T) the Landau levels start to saturate before thermal
broadening sets in. However, we will show that it is not simply
how low in magnetic field we can track the Landau levels, but how
small $1/\nu$ is (where $1/\nu=eB/hn_{s}$), that defines the
trajectory of the extended states as $B\rightarrow$0. In Figure
\ref{phase-diagram} we present a global phase diagram which
reconciles the experimental data from these different material
systems with differing degrees of disorder. This diagram is based
on the global phase diagram (GPD) of Kivelson, Lee, and Zhang
\cite{KLZ} (shown in the inset) modified to highlight the
floating up of the Landau levels in energy. In the GPD (where the
$x$ and $y$-axis are $1/\nu$ and disorder) `floating up' of the
Landau levels in energy as $1/\nu \rightarrow$ 0 is equivalent to
a `floating down' of the quantum Hall liquid-insulator phase
boundary in disorder. To relate data from different material
systems we have normalised for the effective mass (since disorder
$\propto 1/\tau \propto \rho_{xx}/m^*$), and plot the Landau
level trajectories from several studies
\cite{Hanein,Hilke,Shahar,Gusev,Song,Lee,Huang,GD} in the $1/\nu$
-- $1/$disorder plane. The resulting phase diagram in Fig.
\ref{phase-diagram} demonstrates that the data from these
different material systems, with sample disorder varying by three
orders of magnitude, lie approximately on the same curve. This
important result demonstrates that the phase diagrams constructed
by \emph{eight} different studies \emph{are in fact consistent
with each other}, and that it is possible to draw very different
conclusions regarding the fate of the extended states as
$B\rightarrow$ 0 depending on the lowest accessible value of
$1/\nu$. To clarify this further, if we were to delete all the
data below $1/\nu$ = 0.2 in Fig. 4, then \emph{all} studies (even
in systems with no $B$=0 MIT) would appear to show that the Landau
levels do not float up indefinitely as $B \rightarrow$ 0, but
saturate to some finite value. It is only possible to determine
the ultimate fate of the extended states by going to very low
values of $1/\nu$ $(1/\nu \ll 0.1)$.

In summary, we present two significant new findings concerning the
relationship between the $B = 0$ MIT and the quantum Hall effect:
(1) Using high quality n-GaAs systems we have observed that the
extended states that exist in finite magnetic field float
continuously up in energy as $B \rightarrow$ 0, despite an
apparent `metal'-insulator transition at B = 0. (2) Most
importantly, when plotted on a modified form of the Kivelson, Lee
and Zhang global phase diagram, the Landau level trajectories from
this and previous studies in different material systems are in
fact consistent with each other, showing that the extended states
always float up in energy as $B \rightarrow$ 0, irrespective of
whether there is a $B$=0 MIT or not. The phase diagram highlights
the need to go to extremely low values of $1/\nu$ $(1/\nu \ll
0.1)$ in order to determine the ultimate fate of the extended
states as $B \rightarrow$ 0.

This work was funded by the ARC. CEY acknowledges support from
the IPRS scheme; MYS acknowledges an ARC/QEII Fellowship.

\end{multicols}


\begin{references}
\bibitem{note}
Although it is difficult to resolve the transition in the raw data
shown in Fig. 2(f), examination of $\Delta R_{xx} (T)$ reveals a
transition at $n_s^c \sim (8.0 \pm 0.25) \times 10^9$ cm$^{-2}$,
in agreement with Fig. 1.
\bibitem{GD}
To do this we require the resistivity values for each data point,
which were unavailable for Refs. \cite{Glozman,Dultz}.
\end{references}
\end{document}